\documentclass{article}
\usepackage{ijcai26}

\usepackage{times}
\usepackage{soul}
\usepackage{url}
\usepackage[hidelinks]{hyperref}
\usepackage[utf8]{inputenc}
\usepackage[small]{caption}
\usepackage{graphicx}
\usepackage{amsmath}
\usepackage{amsthm}
\usepackage{booktabs}
\usepackage{algorithm}
\usepackage{algorithmic}
\usepackage[switch]{lineno}

\usepackage{multirow}
\usepackage{amssymb}
\usepackage{amsfonts}
\usepackage{booktabs}
\usepackage{xcolor}
\usepackage{graphicx}
\usepackage{makecell}
\usepackage{wrapfig}
\usepackage{diagbox} 

\usepackage{pifont}
\newcommand{\cmark}{\ding{51}}  % ✓
\newcommand{\xmark}{\ding{55}}  % ✗
\usepackage{graphicx}
\usepackage{subcaption}
\linenumbers

\newtheorem{theorem}{Theorem}
\newtheorem{assumption}{Assumption}

\title{PCEvo: Path-Consistent Molecular Representation via Virtual Evolutionary}

\author{
    Author Name
    \affiliations
    Affiliation
    \emails
    email@example.com
}

\author{
Kun Li$^1$
\and
Longtao Hu$^1$\and
Yida Xiong$^1$\and
Jiajun Yu$^2$\and
Hongzhi Zhang$^1$\and
Jiameng Chen$^1$\and \\ 
Xiantao Cai$^1$\and
Jia Wu$^3$\And
Wenbin Hu$^{1,*}$\\
\affiliations
$^1$School of Computer Science, Wuhan University\\
$^2$College of Computer Science and Technology, Zhejiang University\\
$^3$Department of Computing, Macquarie University\\
\emails
\{likun98,hlt\_2003,yidaxiong,zhanghongzhi,	jiameng.chen,caixiantao,hwb\}@whu.edu.cn,
jiajunyu1999@gmail.com,
jia.wu@mq.edu.au
}
\begin{document}
\nolinenumbers % 从此处开始取消行号

\maketitle

\begin{abstract}

Molecular representation learning aims to learn vector embeddings that capture molecular structure and geometry, thereby enabling property prediction and downstream scientific applications. In many AI for science tasks, labeled data are expensive to obtain and therefore limited in availability. Under the few-shot setting, models trained with scarce supervision often learn brittle structure–property relationships, resulting in substantially higher prediction errors and reduced generalization to unseen molecules. To address this limitation, we propose \textbf{PCEvo}, a path-consistent representation method that learns from virtual paths through dynamic structural evolution. PCEvo enumerates multiple chemically feasible edit paths between retrieved similar molecular pairs under topological dependency constraints. It transforms the labels of the two molecules into stepwise supervision along each virtual evolutionary path. It introduces a path-consistency objective that enforces prediction invariance across alternative paths connecting the same two molecules. Comprehensive experiments on the QM9 and MoleculeNet datasets demonstrate that PCEvo substantially improves the few-shot generalization performance of baseline methods.  The code is available at \url{https://anonymous.4open.science/r/PCEvo-4BF2}.

% To address this challenge, we propose PCEvo:,  a path-consistent molecular representation method via dynamic structural evolution. PCEvo: constructs virtual evolutionary edit paths between molecular pairs under topological dependency constraints, decomposes endpoint supervision into edit-wise property increments, and enforces a path consistency constraint to support process-level training over endpoint consistent paths. Moreover, by combining nearest neighbor retrieval with multi-path enumeration, PCEvo: provides a combinatorial data augmentation mechanism that increases the effective training signal without changing the task definition. Experiments on QM9 and MoleculeNet demonstrate that PCEvo: improves generalization and prediction stability in low data settings.

\end{abstract}

\section{Introduction}

% Recent advances in graph neural networks, multimodal contrastive learning, and self-supervised pretraining have substantially advanced molecular representation learning. These techniques have been widely adopted in drug discovery, materials design, and chemical engineering, enabling the construction of expressive molecular representations across multi-scale chemical spaces and supporting tasks such as molecular property prediction and molecular generation.

Recent years have witnessed rapid progress in molecular representation learning driven by graph-based and geometry-aware neural architectures. Message passing networks and 3D geometric models have become mainstream for learning structure-dependent signals, with representative backbones spanning MPNNs \cite{gilmer2017neural}, SchNet \cite{schnet}, DimeNet \cite{gasteiger_dimenet_2020}, and more recent equivariant Transformer families such as Equiformer \cite{equiformer} and ViSNet \cite{visnet}. In parallel, self-supervised and contrastive pretraining has been increasingly adopted to leverage large unlabeled corpora and improve data efficiency, including general graph contrastive objectives \cite{ijcai2024p235} and molecule-specific pretraining paradigms such as GROVER \cite{GROVER}, MolCLR \cite{MolCLR}, and UniMol \cite{UniMol}. These developments are routinely evaluated on established benchmarks such as QM9 \cite{QM9} and MoleculeNet \cite{MoleculeNet}, and are further expanding toward cross-modal alignment between molecular structures and natural language descriptions \cite{wang2025bridging,liu2024moleculargpt}.

\begin{figure}[t!]
    \centering
    \includegraphics[width=0.89\linewidth]{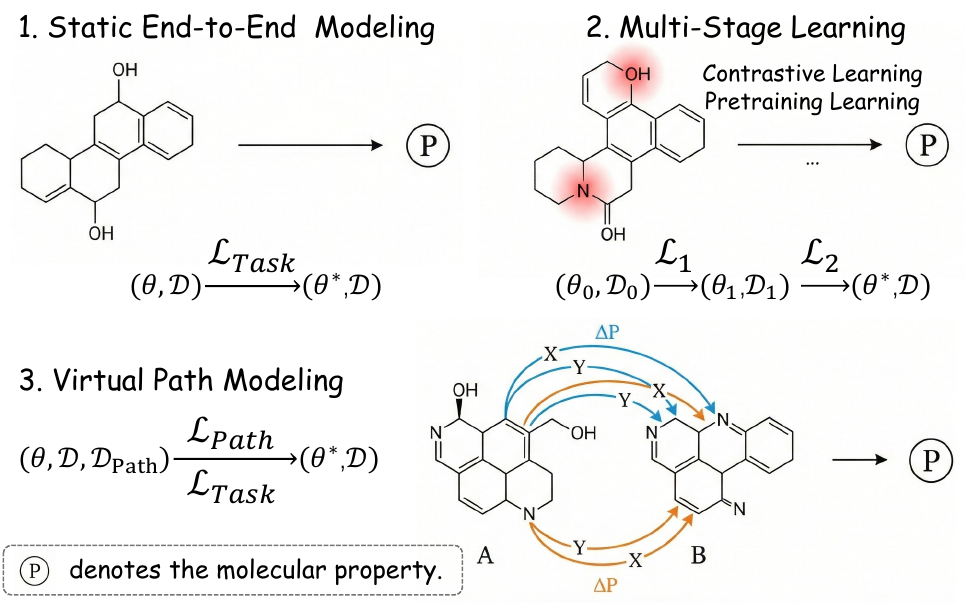}
    \caption{Comparison of molecular representation paradigms: static end-to-end modeling, multi-stage learning, and virtual path modeling. P denotes the molecular property.}
    \label{fig:placeholder}
\end{figure}

% However, most existing methods continue to follow a static modeling paradigm that treats each molecule as an independent graph and learns a direct mapping from a fixed topology to a target property. Although such models perform well in data-rich regimes, their generalization capability degrades significantly in small-sample and sparsely sampled settings. When the coverage of the chemical space is limited, models tend to rely on memorizing structural patterns rather than learning the intrinsic mechanisms that govern property variation, which leads to overfitting and unstable performance. This limitation indicates that static end-point representations fail to capture the process-level information inherent in the formation of molecular structures.

% Despite this progress, the dominant learning paradigm remains largely static: a molecule is treated as a single endpoint sample, and the model is trained to regress properties via direct supervision on sparse labels \cite{empp,zhz} (as shown in Fig.~\ref{fig:placeholder}). 

Despite this progress, the dominant learning paradigm remains largely static: each molecule is treated as a single endpoint sample (i.e., only its final observed structure is used), and the model is trained to regress properties via direct supervision on sparse labels \cite{empp,zhz} (as shown in Fig.~\ref{fig:placeholder}). In few-shot settings, this formulation is prone to overfitting because chemical space is sparsely covered; as a result, models can rely on dataset-specific correlations and frequent motifs, leading to unstable generalization. More fundamentally, most static molecular representations do not explicitly model how properties change under chemically feasible local edits to molecular structure \cite{Mevon}. In practical drug discovery, where labeled assays are costly and task distributions are sparse, capturing such regularities can improve the robustness of learned representations under the few-shot setting.

\begin{figure}[t!]
    \centering
    \includegraphics[width=0.89\linewidth]{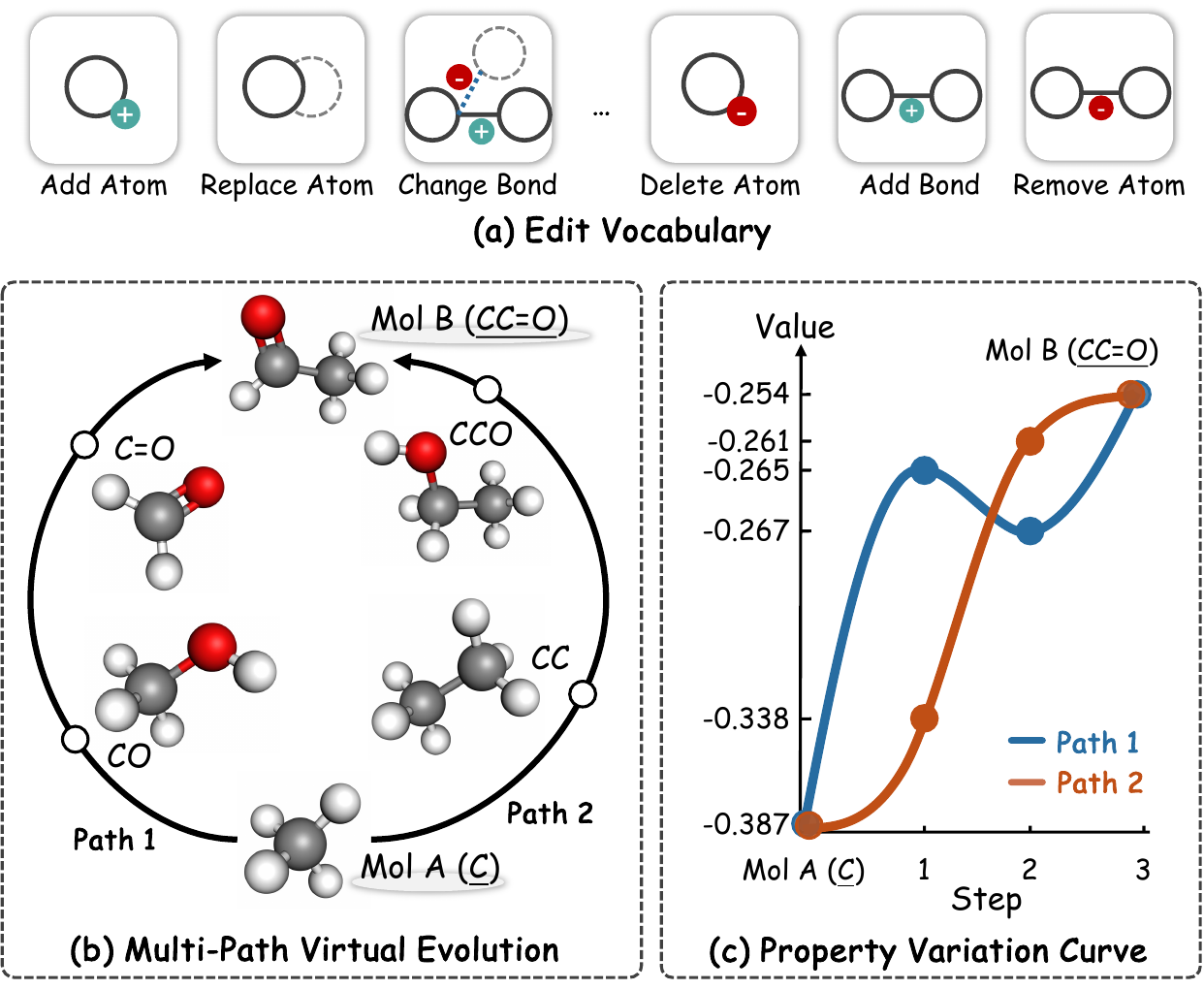}
    \caption{Virtual evolutionary editing uses the HOMO energy from the QM9 dataset to illustrate two evolution paths between a pair of molecules. Despite exhibiting different intermediate fluctuations, both paths inevitably converge to the same final HOMO value, highlighting the path‐independent nature of the property.}
    \label{fig:mov}
\end{figure}

From a chemical perspective, molecules are discrete combinatorial objects whose structures can be naturally described as graphs and encoded by rule-based line notations \cite{gilmer2017neural,weininger1988smiles}. More importantly, property changes in medicinal chemistry are often analyzed through small, chemically valid transformations, where a minimal structural modification induces a measurable shift in activity or physicochemical profiles \cite{ishikura2025amino,hussain2010mmp}. These observations motivate a virtual evolutionary path view of the structure--property relationship (Fig.~\ref{fig:mov}) \cite{maggiora2014chemical,stumpfe2012exploring}. By treating minimal edit operations, such as atom substitution, local connectivity adjustment, and functional group modification, as fundamental units \cite{walz2025macrocyclization,hussain2010computationally}, we can construct virtual evolutionary paths connecting different molecular states \cite{jensen2019graph}. Modeling property variation along such paths enables the models to leverage incremental structural signals that are often underutilized when training solely on static endpoint samples.

% From a chemical perspective, molecules are discrete combinatorial objects whose structures can be naturally described as graphs and encoded by rule-based line notations \cite{weininger1988smiles}. More importantly, property changes in medicinal chemistry are often analyzed through small, chemically valid transformations, where a minimal structural modification induces a measurable shift in activity or physicochemical profiles \cite{hussain2010mmp}. These observations motivate a process-oriented view of the structure--property relationship (Fig.~\ref{fig:mov}). By treating minimal edit operations, such as atom substitution, local connectivity adjustment, and functional group modification, as fundamental units, we can construct virtual evolutionary paths connecting different molecular states. Modeling property variation along such paths allows the learner to exploit incremental structural signals that are largely underutilized when training solely on static end-point graphs.

% Path-Consistent Molecular Representation via Virtual Evolutionary

% Based on this perspective, we propose a \textbf{p}rocess-\textbf{a}ware molecular representation framework via dynamic \textbf{s}tructural \textbf{e}volution, referred to as \textbf{PASE}. In PASE, each molecule is treated as a terminal state that can be reached through multiple virtual edit paths, and the responses of minimal edits are learned to decompose complex structural variations into additive components. Extensive experiments on QM9 and MoleculeNet under few-shot settings demonstrate that PASE consistently improves predictive performance across diverse backbone architectures, achieving 18--45\% reductions in MAE. Our contributions are summarized as follows:

Based on this perspective, we propose \textbf{PCEvo}, a \textbf{p}ath-\textbf{c}onsistent representation method via virtual \textbf{evo}lutionary. PCEvo does not treat molecules as isolated, static samples; instead, it represents the structural differences between two similar molecules as an evolutionary sequence of chemically feasible minimal edit operations. PCEvo further converts the labels of the two endpoint molecules into stepwise supervision distributed along the path, enabling the model to learn how incremental structural changes drive property variations during training, rather than performing the regression only at the endpoints. Importantly, there are typically multiple feasible editing paths connecting the same molecular pair; yet, the cumulative property change should remain consistent regardless of which path is taken. Leveraging this principle, PCEvo introduces a path-consistency objective that aligns the accumulated predictions across alternative feasible paths. This objective integrates seamlessly with standard static supervised training, thereby improving generalization in few-shot settings. We evaluate PCEvo on QM9 and MoleculeNet by integrating it into a range of representative backbone methods, thereby systematically demonstrating its generality and effectiveness; results show that PCEvo consistently reduces prediction error under few-shot settings on QM9 and achieves state-of-the-art (SOTA) performance on the three MoleculeNet regression tasks under the standard split, while maintaining reliable gains in both accuracy and stability when labeled data are limited.

% We evaluate PCEvo on QM9 and MoleculeNet datasets by integrating it into a range of representative backbone methods, thereby systematically demonstrating its generality and effectiveness. Under few-shot settings, PCEvo achieves state-of-the-art (SOTA) performance compared with strong baselines. Our contributions are summarized as follows:

% PCEvo enumerates multiple chemically feasible edit paths between retrieved similar molecular pairs under topological dependency constraints. It transforms the labels of the two molecules into stepwise supervision along each path. We introduce a path-consistency objective that enforces prediction invariance across alternative paths connecting the same two molecules. Comprehensive experiments on the QM9 and MoleculeNet datasets demonstrate that PCEvo substantially improves the few-shot generalization performance of baseline methods. Our contributions are summarized as follows:

\begin{itemize}
    % \item We propose a process-aware molecular representation framework that models molecules through virtual structural evolution paths rather than static end-point samples, and represents molecular variation by learning additive edit-level responses in property space.

    \item We propose PCEvo, a path-consistent molecular representation method that learns from virtual evolutionary edit paths between similar molecules, transforming sparse end-point supervision into stepwise training signals along chemically feasible paths.

    % \item PCEvo is designed to integrate seamlessly with existing molecular backbones: the backbone performs representation learning and property prediction, while PCEvo adds path-level supervision and a path-consistency objective during training to improve robustness under limited labels.

    % \item A path-integration-based generalization analysis proves that edit-level decomposition yields a smaller error bound by simultaneously increasing effective sample size and constraining hypothesis complexity.
    % \item Extensive experiments on QM9 and MoleculeNet under few-shot settings show that PCEvo consistently reduces MAE by 18-45\% across diverse backbone methods.
    % \item We evaluate PCEvo on QM9 and MoleculeNet datasets by integrating it into a range of representative backbone methods, thereby systematically demonstrating its generality and effectiveness. Under few-shot settings, PCEvo achieves state-of-the-art (SOTA) performance compared with strong baselines
    \item Extensive experiments on QM9 and MoleculeNet under few-shot settings show that PCEvo consistently reduces prediction error and improves predictive stability, achieving SOTA performance compared with strong baselines.
\end{itemize}

\section{Related Works}
\label{sec:related_work}

Mainstream end-to-end molecular representation methods can be broadly grouped into 2D topology-based GNNs \cite{gat,gcn,GraphSAGE} and 3D geometry-aware models \cite{batatia2025design,visnet,musaelian2023learning,equiformer,schutt2021equivariant}. Attention-enhanced encoders have also been explored to improve long-range dependency modeling and the expressivity--efficiency trade-off \cite{qin2025moleculeformer,gotennet}. Despite continuous progress in backbone design, most models are still trained under a static end-to-end regression paradigm, which tends to be sample-inefficient and unstable when supervision is extremely limited or chemical space coverage is insufficient.

% To improve transferability and robustness in low-resource settings, multi-stage training paradigms, most notably pretrain-finetune, self-supervised/contrastive learning, have become widely adopted \cite{aldossary2024silico,UniMol,sun2022does,Hu2020pretrain}. Representative directions include unsupervised structural statistics \cite{NGRAM}, large-scale molecular pretraining \cite{GROVER,Zhou2023unimol,GEM}, and contrastive representation learning \cite{li2025contrastive,ijcai2024p234,ijcai2025p303,MolCLR}. Nevertheless, these approaches remain largely endpoint-centric and do not explicitly model how structural changes translate into property variations.

To alleviate the sample inefficiency, recent studies increasingly adopt multi-stage learning paradigms that leverage large-scale unlabeled data and auxiliary supervision to enhance representation transferability under limited training data \cite{aldossary2024silico,UniMol,sun2022does}. Typical methods include representation learning based on unsupervised structural statistics \cite{NGRAM}, pretraining followed by task-specific fine-tuning \cite{UniMol,GEM,GROVER}, and self-supervised or contrastive objectives that regularize representations through augmentation-invariant constraints \cite{li2025contrastive,ijcai2025p303,ijcai2024p234,MolCLR}. However, most methods still treat molecules as independent static samples, largely ignoring cross-sample structural relatedness and transferable regularities.

\begin{figure*}[t!]
    \centering
    \includegraphics[width=0.9\linewidth]{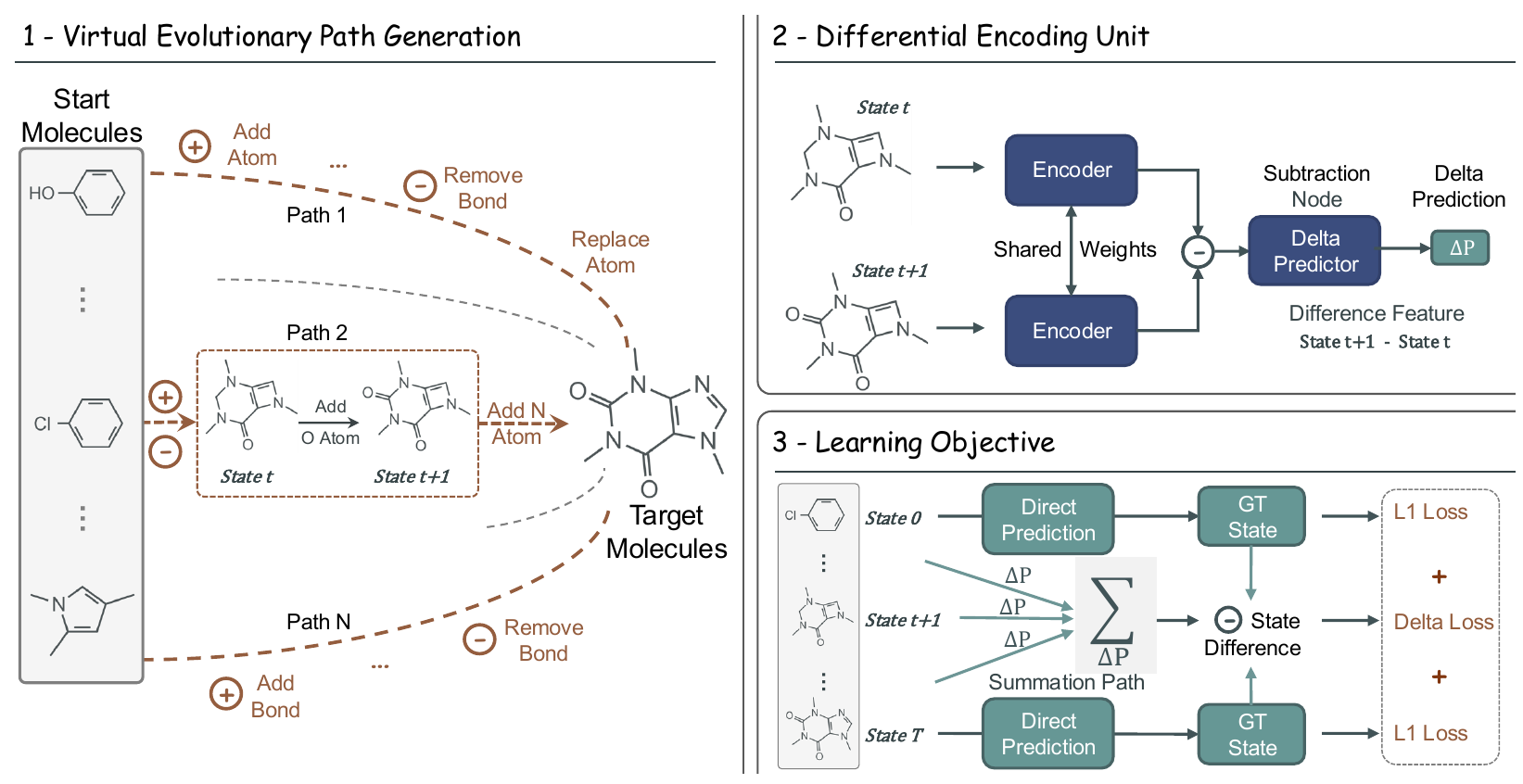}
    \caption{Overview of PCEvo method. Our method constructs virtual evolutionary paths between molecular states by identifying a sequence of valid graph edit operations. These discrete operations are then mapped into a continuous representation space to guide the learning of chemically interpretable molecular representations.}
    \label{fig:framework}
\end{figure*}

\section{Methods}

\subsection{Overview}

To address the generalization limitations of static end-to-end learning in few-shot scenarios, as illustrated in Fig.~\ref{fig:framework}, we propose a path-consistent molecular representation method, PCEvo. Deviating from conventional static structural mapping paradigms, PCEvo decomposes complex structure-property correlations into a sequence of learnable, cumulative edit steps. By imposing these path constraints, PCEvo could capture the intrinsic logic underlying molecular evolution.

\subsection{Virtual Evolutionary Path Generation}
\label{sec:path_generation}

To transform static molecular graphs into dynamic evolutionary paths, we propose a virtual evolutionary path generation pipeline. This pipeline converts a standard dataset $\mathcal{D} = \{(G_i, y_i)\}_{i=1}^N$ into an evolutionary dataset $\mathcal{D}_{\mathrm{evo}} = \{(G_s, G_t, \tau) \mid \tau \in \mathcal{T}(G_s, G_t)\}$. For each molecular pair $(G_s, G_t)$, $\mathcal{T}(G_s, G_t)$ denotes the set of all valid evolutionary paths connecting the source molecule $G_s$ to the target molecule $G_t$. Each path $\tau \in \mathcal{T}(G_s, G_t)$ is a sequence of edit operations. Explicitly modeling multiple valid paths for each molecular pair allows the PCEvo to account for the intrinsic diversity of structural transformations, facilitating the learning of property changes that are invariant to the ordering of operations.

% This process converts a standard dataset $\mathcal{D} = \{G_i, y_i\}_{i=1}^N$ into a set of evolutionary triplets $\mathcal{D}_{evo} = {\{(G_s, G_t, \tau_1), (G_s, G_t, \tau_k2), ..., (G_s, G_t, \tau_k)\}}$, where $\tau_k$ represents a valid sequence of edit operations transforming a source molecule $G_s$ to a target $G_t$.

\subsubsection{Structural Neighbor Retrieval}
Constructing evolutionary paths between arbitrary pairs of molecules is chemically implausible and computationally inefficient. To model realistic structural optimizations, we restrict the source molecules to the chemical neighborhood of the target.

For each target molecule $G_t$ in the training or validation set, we query a candidate pool $\mathcal{G}_{s}$ to identify the top-$\text{K}$ nearest neighbors. We utilize the Tanimoto similarity $f_{\textit{Tanimoto}}(G_s,G_t)$ coefficient \cite{tanimoto} based on extended-connectivity fingerprints as the metric:
\begin{equation}
    \mathcal{N}(G_t) = \text{Top}_\text{K} \left( \{ G_s \in \mathcal{G}_{src} \mid G_s \neq G_t \}, f_{\textit{Tanimoto}} \right).
\end{equation}

This retrieval-based pairing strategy ensures that $G_s$ and $G_t$ share a significant structural scaffold, thereby defining a chemically meaningful optimization landscape.

\subsubsection{Fundamental Path Operation Edit Units}
Identifying the minimal set of edit operations ensures that the molecular transformation includes only essential structural changes, improving path generation efficiency and facilitating the learning of structure–property relationships. Given a pair $(G_s, G_t)$, we seek the minimal set of edit operations $\mathcal{S}$ required to transform $G_s$ into $G_t$. We formulate this as a maximum common subgraph (MCS) problem \cite{fuchs2025fast,mcs1,mcs2}.

We establish an atom-level alignment $\pi: V_s \to V_t$ by minimizing the matching cost matrix defined by atomic numbers and valence constraints. Based on the atom-level alignment $\pi$, the symmetric difference between the two graphs is decomposed into a set of atomic operations. The definitions of these operations are detailed in Table~\ref{tab:graph_ops_def}. Specifically, unmapped atoms in the source molecule $G_s$ correspond to \texttt{REMOVE\_ATOM} operations, while unmapped atoms in the target molecule $G_t$ correspond to \texttt{ADD\_ATOM} operations. At the bond level, discrepancies between mapped atom pairs $(u, v)$ in $G_s$ and $(\pi(u), \pi(v))$ in $G_t$ give rise to \texttt{ADD\_BOND}, \texttt{REMOVE\_BOND}, or \texttt{CHANGE\_BOND} operations. 

% Any mismatches in atom types are handled by \texttt{REPLACE\_ATOM} operations.

Applying this procedure yields an unordered set of operations $\mathcal{S} = \{o_1, \dots, o_M\}$, which constitutes the minimal set of edits required to transform $G_s$ into $G_t$.

\subsubsection{Topologically Constrained Path Sampling}

While the set $\mathcal{S}$ specifies the structural changes, it does not determine the order in which they should be applied. Performing the operations in an arbitrary sequence may violate topological constraints, such as adding a bond to an atom that does not yet exist. To resolve this, we construct a directed acyclic graph, termed the dependency graph: $\mathcal{G}_{dep} = (\mathcal{S}, \mathcal{E}_{dep})$. An edge $o_i \to o_j$ exists if operation $o_i$ is a topological prerequisite for $o_j$. The core constraints include:
\begin{itemize}
    \item Existence precondition: An \texttt{ADD\_BOND}$(u, v)$ operation depends on the \texttt{ADD\_ATOM} operations that create nodes $u$ and $v$.
    \item Deletion latency: A \texttt{REMOVE\_ATOM}$(u)$ operation must succeed all bond removals or modifications involving node $u$.
\end{itemize}

Any linear extension (topological sort) of $\mathcal{G}_{dep}$ constitutes a valid path $\tau$. Importantly, a single edit set $\mathcal{S}$ often allows exponentially many valid permutations. We leverage this by randomly sampling up to $P_{max}$ distinct topological sorts for each molecular pair. This serves as a powerful combinatorial data augmentation strategy: the model learns to associate the same net property change $\Delta y$ with multiple distinct paths $\{\tau_1, \dots, \tau_{P_{max}}\}$, thereby enforcing path-independence of the physical property and improving generalization. Each path $\tau \in \mathcal{T}(G_s, G_t)$ is a sequence of edit operations, and its length may vary across different paths.

\begin{algorithm}[t]
\caption{Virtual Evolutionary Path Generation}
\label{alg:path_gen}
\begin{algorithmic}[1]
\REQUIRE Standard dataset $\mathcal{D}$, Number of neighbors $\text{K}$, Max paths $P_{max}$
\ENSURE Evolutionary dataset $\mathcal{D}_{\mathrm{evo}}$
\STATE Initialize $\mathcal{D}_{\mathrm{evo}} \leftarrow \emptyset$
\FORALL{$G_t \in \mathcal{D}$}
    \STATE \text{Candidate pool} $\mathcal{G}_{s} \leftarrow (G_t, \mathcal{G}_{s}, K)$
    \FORALL{$G_s \in \mathcal{G}_{s} $}
        \STATE $\pi \leftarrow \textsc{Alignment}(G_s, G_t)$
        \STATE $\mathcal{S} \leftarrow \textsc{ExtractMinimalEdits}(G_s, G_t, \pi)$
        \STATE $\mathcal{G}_{dep} \leftarrow \textsc{BuildDependencyGraph}(\mathcal{S})$
        \STATE $\mathcal{T}_{pair} \leftarrow \emptyset$
        \WHILE{$|\mathcal{T}_{pair}| < P_{max}$}
            \STATE $\tau \leftarrow \textsc{Sampling}(\mathcal{G}_{dep})$
            \STATE $\mathcal{T}_{pair} \leftarrow \mathcal{T}_{pair} \cup \{\tau\}$
        \ENDWHILE
        \FORALL{$\tau \in \mathcal{T}_{pair}$}
            \STATE $\mathcal{D}_{\mathrm{evo}} \leftarrow \mathcal{D}_{\mathrm{evo}} \cup \{(G_s, G_t, \tau)\}$
        \ENDFOR
    \ENDFOR
\ENDFOR
\RETURN $\mathcal{D}_{\mathrm{evo}}$
\end{algorithmic}
\end{algorithm}

\subsection{Differential Evolutionary Path Encoder}
\label{sec:diff_encoder}

To represent the feature changes induced by each virtual edit operation along a molecular structural evolution path, a differential evolutionary path encoder is proposed. Unlike the standard paradigm that encodes molecules as isolated static data points, this module explicitly models the incremental changes between consecutive intermediate states along the virtual path $\tau$. 

Let $G_s^{(t)}$ denote the intermediate molecular state after the $t$-th edit operation in the path, with the boundary conditions $G_s^{(0)} = G_s$  and $G_s^{(T)} = G_t$, where $T$ is the total path length.

To bridge the discrete symbolic space of chemical edits and the continuous latent space, we first define a bijective mapping $\psi: \mathcal{O} \to \mathbb{R}^D$. The $t$-th edit operation $o_t$, which triggers the transition $G_s^{(t)} \to G_s^{(t+1)}$, is vectorized via concatenated encodings:
\begin{equation}
    \mathbf{x}_t = \psi(o_t) = [\mathbf{e}_{type} \parallel \mathbf{e}_{atom} \parallel \mathbf{e}_{bond} \parallel \mathbf{e}_{pos}],
\end{equation}
\noindent where components encode the operation type, atomic semantics, bond attributes, and positional indices, respectively. This vectorization provides the initial condition for the specific structural transformation at step $t$.

Then, we employ a shared-weight molecular encoder $f_\theta(\cdot)$ to project the adjacent states $G_s^{(t)}$ and $G_s^{(t+1)}$ into a continuous latent manifold:
\begin{equation}
    \mathbf{h}_t = f_\theta(G_s^{(t)}), \quad \mathbf{h}_{t+1} = f_\theta(G_s^{(t+1)}).
\end{equation}

% Importantly, our method is architecture-agnostic, meaning that the encoder $f_\theta$ is not limited to graph neural networks (GNNs) and can be instantiated as any representation model capable of extracting features from molecular data (e.g., Transformer, CNN, or geometric networks). Parameter sharing ensures that the feature extraction logic remains consistent across different steps $t$ along the evolutionary path.

% To isolate the net physical impact induced by the $t$-th structural modification, we compute a difference feature vector via an element-wise subtraction node:
% \begin{equation}
%     \mathbf{d}_{t} = \mathbf{h}_{t+1} - \mathbf{h}_t.
% \end{equation}
% This differential vector $\mathbf{d}_{t}$ captures the relative structural variation resulting from the virtual edit. It is subsequently fed into the delta predictor $\phi_\psi(\cdot)$ (implemented as a multi-layer perceptron) to estimate the scalar property change:
% \begin{equation}
%     \Delta \hat{P}_t = \phi_\psi(\mathbf{d}_{t}).
%     \label{eq:delta}
% \end{equation}
% By focusing on the relative delta rather than absolute values, this design aligns with the physical principle of additivity, allowing the model to explicitly learn how specific structural edits influence property changes.

Our method PCEvo is architecture-agnostic and shared across all steps to ensure consistent feature extraction along the path. For each virtual edit, we model its incremental effect by differencing adjacent representations and predicting the corresponding property change:
\begin{equation}
\mathbf{d}_{t}=\mathbf{h}_{t+1}-\mathbf{h}_{t},
\qquad
\Delta \hat{P}_t=\phi_\psi(\mathbf{d}_{t}),
\label{eq:delta}
\end{equation}
which learns edit-wise contributions in an additive manner and $\phi_\psi(\cdot)$ is non-linear delta predictor.

\begin{table}[t!]
    \centering
    \caption{Elementary Graph Edit Operations.}
    \label{tab:graph_ops_def}
    \resizebox{0.95\linewidth}{!}{
    \begin{tabular}{ll}
        \toprule
        Operation & Description \\
        \midrule
        \texttt{ADD/REM/REP\_ATOM} & Adds, removes, or replaces an atom node. \\
        \texttt{ADD/REM/CHG\_BOND} & Adds, removes, or modifies a bond edge. \\
        \bottomrule
    \end{tabular}
    }
\end{table}

\subsection{Learning Objective}

Our optimization jointly anchors absolute property prediction in the global chemical space and enforces summation consistency of property variations along virtual evolutionary paths. 
The model consists of a shared-weight backbone encoder $f_\theta(\cdot)$ and a non-linear delta predictor $\phi_\psi(\cdot)$. 
Given a labeled molecular pair $(G_s,G_t)$ with properties $(y_s,y_t)$ and a feasible edit path $\tau=\big(G_s^{(0)},\ldots,G_s^{(T)}\big)$ satisfying $G_s^{(0)}=G_s$ and $G_s^{(T)}=G_t$, we denote the latent representation at step $t$ as
\begin{equation}
    h_t = f_\theta\!\left(G_s^{(t)}\right),\qquad t=0,1,\ldots,T.
\end{equation}

\paragraph{Static Property Loss.}
We supervise absolute predictions at the observed endpoints:
\begin{equation}
\mathcal{L}_{\text{static}}
=
\left\|f_\theta(G) - y\right\|_1
\end{equation}

\paragraph{Path Summation Consistency Loss.}
We require that the cumulative sum of predicted stepwise property increments along the path matches the ground-truth endpoint difference:
\begin{equation}
\mathcal{L}_{\text{cons}}
=
\left\|
\widehat{\Delta P}(\tau) - (y_t-y_s)
\right\|_1,
\label{eq:loss_cons}
\end{equation}
% \begin{equation}
% \widehat{\Delta P}(\tau)
% =
% \sum_{t=0}^{T-1}\phi_\psi\!\left(h_{t+1}-h_t\right).
% \label{eq:deltap_hat}
% \end{equation}
\begin{equation}
\widehat{\Delta P}(\tau)
=
\sum_{t=0}^{T-1}\Delta \hat{P}_t
=
\sum_{t=0}^{T-1}\phi_\psi\!\left(\mathbf{h}_{t+1}-\mathbf{h}_t\right).
\label{eq:deltap_hat}
\end{equation}

\noindent where the non-linear predictor $\phi_\psi(\cdot)$ ensures the summation does not mathematically collapse into a simple difference of endpoints. This forces the gradient signal to propagate through every intermediate state, compelling the model to learn the specific physical contribution of each elementary edit operation. The final objective is $\mathcal{L} = \mathcal{L}_{\text{static}} + \mathcal{L}_{\text{cons}}$.

% \newpage

\begin{table*}[t!]
  \centering
  \caption{Overall experiments on QM9 dataset for \text{HOMO} and \text{LUMO} (units: eV) properties under few-shot settings. \textbf{Origin} denotes the baseline backbone, and \textbf{Improve.}(\%)  indicates the performance gain.}
  \label{tab:qm9_results}
  \resizebox{\textwidth}{!}{%
    \begin{tabular}{cr|ccc|ccc|ccc|ccc}
    \toprule
    % \multicolumn{2}{c|}{\multirow{3}{*}{{\textbf{Methods}}}} & \multicolumn{6}{c|}{$\epsilon_{\text{HOMO}}$ } & \multicolumn{6}{c}{$\epsilon_{\text{LUMO}}$} \\
    \cmidrule(lr){3-8} \cmidrule(lr){9-14}
    \multicolumn{2}{c|}{\multirow{3}{*}{{\textbf{Methods}}}} & \multicolumn{3}{c|}{\text{HOMO} (100-shot)} & \multicolumn{3}{c|}{\text{HOMO} (1000-shot)} & \multicolumn{3}{c|}{\text{LUMO} (100-shot)} & \multicolumn{3}{c}{\text{LUMO} (1000-shot)} \\
    \cmidrule(lr){3-5} \cmidrule(lr){6-8} \cmidrule(lr){9-11} \cmidrule(lr){12-14}
    & & \textbf{MAE} $\downarrow$ & \textbf{MSE} $\downarrow$ & \textbf{PCC} $\uparrow$ & \textbf{MAE} $\downarrow$ & \textbf{MSE} $\downarrow$ & \textbf{PCC} $\uparrow$ & \textbf{MAE} $\downarrow$ & \textbf{MSE} $\downarrow$ & \textbf{PCC} $\uparrow$ & \textbf{MAE} $\downarrow$ & \textbf{MSE} $\downarrow$ & \textbf{PCC} $\uparrow$ \\
    
    \midrule
    % \multicolumn{2}{c|}{Moleculeformer \cite{qin2025moleculeformer} } & 0.4688 & 0.4917 & 0.3413 & 0.3617 & 0.2580 & 0.6336 & 0.8985 & 1.2785& 0.4202 & 0.5642 & 0.5420 & 0.8192\\
    % \multicolumn{2}{c|}{Moleculeformer \cite{qin2025moleculeformer} } & 0.4688 & 0.4917 & 0.3413 & 0.3617 & 0.2580 & 0.6336 & 0.8985 & 1.2785& 0.4202 & 0.5642 & 0.5420 & 0.8192\\
    % \multicolumn{2}{c|}{Moleculeformer \cite{qin2025moleculeformer} } & 0.4688 & 0.4917 & 0.3413 & 0.3617 & 0.2580 & 0.6336 & 0.8985 & 1.2785& 0.4202 & 0.5642 & 0.5420 & 0.8192\\

    \multicolumn{2}{c|}{GCN \cite{gcn} } &     0.5031 & 0.4298 & -0.2760 & 0.4318 & 0.3393 & 0.3345 & 0.7651 & 0.8399 & 0.6577 & 0.6330 & 0.6617 & 0.7859  \\

    \multicolumn{2}{c|}{GraphSAGE \cite{GraphSAGE} } &     0.5121 & 0.4523 & -0.1318 & 0.4759 & 0.4077 & -0.0035 & 0.7618 & 0.8946 & 0.6397 & 0.7134 & 0.7961 & 0.7322 \\
    
    \multicolumn{2}{c|}{GAT \cite{gat} } &     0.5016 & 0.4290 & -0.2948 & 0.7613 & 0.8418 & 0.1641 & 0.7682 & 0.8763 & 0.6408 & 0.7234 & 0.8132 & 0.7238 \\

    \multicolumn{2}{c|}{GIN \cite{gin} } &     0.4928 & 0.4190 & -0.1601 & 0.3529 & 0.2176 & 0.6347 & 0.6768 & 0.7474 & 0.7225 & 0.6001 & 0.6568 & 0.8257 \\

    \multicolumn{2}{c|}{EMPP \cite{empp} } & 0.4058 & 0.2872 & 0.5820 & 0.2733 & 0.1244 & 0.8115 & 0.7328 & 0.7804 & 0.6983 & 0.2713 & 0.1223 & 0.8150 \\
    \multicolumn{2}{c|}{EMPP$^*$ \cite{empp} } & 0.7520 & 0.8391 & 0.0587 & 2.0454 & 4.6518 & -0.2126 & 0.9696 & 1.4393 & 0.6016 & 1.0356 & 1.5991 & 0.6211 \\
    \multicolumn{2}{c|}{Moleculeformer \cite{qin2025moleculeformer} } &     0.3353 & 0.2165 & 0.6966 & 0.2710 & 0.1274 & 0.8068 & 0.8985 & 1.2785& 0.4202 & 0.5642 & 0.5420 & 0.8192\\
    \multicolumn{2}{c|}{GotenNet \cite{gotennet} } & 0.5565 & 0.6819 & -0.0591 & 0.3743 & 0.2860 & 0.5950 & 0.8360 & 1.1393 & 0.5379 & 0.4194 & 0.2938 & 0.9067\\
    
    \midrule

    % \hline
    
    \multirow{3}{*}{\makecell{SchNet \\\cite{schnet}} }
    & \textit{+PCEvo} & \textbf{0.3330} & \textbf{0.1853} & \textbf{0.7694} & \textbf{0.2057} & \textbf{0.0758} & \textbf{0.8954} & \textbf{0.5275} & \textbf{0.5253} & \textbf{0.8494} & \textbf{0.2269} & \textbf{0.0924} & \textbf{0.9712} \\
    & \textit{Origin.} & 0.4688 & 0.4917 & 0.3413 & 0.3617 & 0.2580 & 0.6336 & 0.6159 & 0.6006 & 0.8050 & 0.2944 & 0.1551 & 0.9514 \\
    & \textit{Improve.} & \textcolor{red}{28.97\%} & \textcolor{red}{62.30\%} & \textcolor{red}{125.41\%} & \textcolor{red}{43.13\%} & \textcolor{red}{70.61\%} & \textcolor{red}{41.31\%} & \textcolor{red}{14.35\%} & \textcolor{red}{12.54\%} & \textcolor{red}{5.52\%} & \textcolor{red}{22.94\%} & \textcolor{red}{40.45\%} & \textcolor{red}{2.08\%} \\
    \midrule
    
    \multirow{3}{*}{\makecell{DimeNet \\\cite{gasteiger_dimenet_2020}} }
    & \textit{+PCEvo}  & \textbf{0.5502} & \textbf{0.4888} & \textbf{0.3362} & \textbf{0.3711} & \textbf{0.2373} & \textbf{0.6824} & \textbf{0.6955} & 0.8305 & \textbf{0.7117} & \textbf{0.5553} & \textbf{0.5210} & \textbf{0.8242} \\
    & \textit{Origin.} & 0.6162 & 0.6871 & 0.1406 & 0.4392 & 0.3322 & 0.4835 & 0.7173 & \textbf{0.7877} & 0.6814 & 0.5642 & 0.5420 & 0.8192 \\
    & \textit{Improve.}  & \textcolor{red}{10.71\%} & \textcolor{red}{28.86\%} & \textcolor{red}{139.04\%} & \textcolor{red}{15.50\%} & \textcolor{red}{28.55\%} & \textcolor{red}{41.13\%} & \textcolor{red}{3.04\%} & \textcolor[rgb]{0.5,0.5,0.5}{-5.43\%} & \textcolor{red}{4.45\%} & \textcolor{red}{1.58\%} & \textcolor{red}{3.88\%} & \textcolor{red}{0.61\%} \\
    \midrule
    
    \multirow{3}{*}{\makecell{Equiformer \\\cite{equiformer}}}
    & \textit{+PCEvo}  & \textbf{0.3971} & \textbf{0.3152} & \textbf{0.5319} & \textbf{0.2292} & \textbf{0.0963} & \textbf{0.8637} & \textbf{0.5463} & \textbf{0.5210} & \textbf{0.8055} & \textbf{0.2592} & \textbf{0.1292} & \textbf{0.9593} \\
    & \textit{Origin.} & 0.5189 & 0.5251 & 0.0726 & 0.3098 & 0.1698 & 0.7366 & 0.8261 & 1.0350 & 0.5609 & 0.4264 & 0.2920 & 0.9057 \\
    & \textit{Improve.}  & \textcolor{red}{23.46\%} & \textcolor{red}{39.98\%} & \textcolor{red}{631.98\%} & \textcolor{red}{26.02\%} & \textcolor{red}{43.29\%} & \textcolor{red}{17.25\%} & \textcolor{red}{33.87\%} & \textcolor{red}{49.66\%} & \textcolor{red}{43.60\%} & \textcolor{red}{39.19\%} & \textcolor{red}{55.72\%} & \textcolor{red}{5.91\%} \\
    \midrule
    
     \multirow{3}{*}{\makecell{ViSNet \\ \cite{visnet}}} 
    & \textit{+PCEvo}  & \textbf{0.4570} & \textbf{0.4725} & \textbf{0.4609} & \textbf{0.2346} & \textbf{0.1001} & \textbf{0.8696} & \textbf{0.4073} & \textbf{0.2905} & \textbf{0.9035} & \textbf{0.2490} & \textbf{0.1178} & \textbf{0.9635} \\
    & \textit{Origin.} & 0.5889 & 0.7685 & -0.1158 & 0.3181 & 0.2159 & 0.7107 & 0.4852 & 0.3693 & 0.8791 & 0.2585 & 0.1297 & 0.9603 \\
    & \textit{Improve.} & \textcolor{red}{22.39\%} & \textcolor{red}{38.52\%} & \textcolor{red}{497.95\%} & \textcolor{red}{26.25\%} & \textcolor{red}{53.62\%} & \textcolor{red}{22.36\%} & \textcolor{red}{16.04\%} & \textcolor{red}{21.33\%} & \textcolor{red}{2.78\%} & \textcolor{red}{3.70\%} & \textcolor{red}{9.19\%} & \textcolor{red}{0.32\%} \\
    \bottomrule
    \multicolumn{14}{l}{\footnotesize $*$ denotes EMPP trained with an auxiliary objective that additionally predicts masked atomic coordinates.} \\
    \end{tabular}%
  }
\end{table*}

\begin{table}[t!]
\centering
\caption{Overall experiments on MoleculeNet under standard data splits. RMSE is reported for each task, and \textbf{Avg.} denotes the mean RMSE over all three tasks.}
\label{tab:MoleculeNetStandard}
\resizebox{\columnwidth}{!}{ % 自动缩放以适应单栏宽度
\begin{tabular}{l cccc}
\toprule
\textbf{Methods} & ESOL & FreeSolv & Lipophilicity & \textbf{Avg.} \\
\midrule
GCN \cite{gcn} & 1.431 & 2.870 & \textbf{0.712} & 1.671 \\
GIN \cite{gin} & 1.452 & 2.765 & 0.850 & 1.689 \\
N-GRAM \cite{NGRAM} & 1.100 & 2.510 & 0.880 & 1.497 \\
Hu et al. \cite{Huat} & 1.100 & 2.764 & \underline{0.739} & 1.534 \\
GROVER \cite{GROVER} & 1.423 & 2.947 & 0.823 & 1.731 \\
MolCLR \cite{MolCLR} & 1.113 & 2.301 & 0.789 & 1.401 \\
Equiformer \cite{equiformer} & 1.037 & 2.672 & 0.989 & 1.566 \\
\midrule
SchNet \cite{schnet} & \textbf{0.912} & 2.255 & 0.923 & 1.363 \\
DimeNet \cite{gasteiger_dimenet_2020} & 1.135 & 2.589 & 1.043 & 1.589 \\
ViSNet \cite{visnet} & 1.049 & 3.158 & 0.944 & 1.717 \\

\midrule
$\text{PCEvo}_{\;\text{SchNet}}$ (Ours) & 1.060 & \underline{1.958} & 0.762 & \textbf{1.260} \\
$\text{PCEvo}_{\;\text{DimeNet}}$ (Ours)& 1.117 & \textbf{1.843} & 0.925 & \underline{1.295} \\
$\text{PCEvo}_{\;\text{ViSNet}}$ (Ours)& \underline{0.969} & 2.539 & 0.798 & 1.436 \\
\bottomrule
\end{tabular}
}
\end{table}

\begin{table}[t!]
\centering
% \caption{RMSE comparison on ESOL, FreeSolv, and Lipophilicity under different train/test splits. \textbf{R} means the training sample ratio, \textbf{Origin} denotes the baseline backbone, and \textbf{\textit{Imp.}} indicates the performance gain.}
\caption{Overall experiments on MoleculeNet under different train/test splits. RMSE is reported for each task, \textbf{R} denotes the training sample ratio.}
\label{tab:MoleculeNetFewShot}
\setlength{\tabcolsep}{2pt} 
\resizebox{\columnwidth}{!}{ % 自动缩放以适应单栏宽度
\scriptsize 
\begin{tabular}{cr|ccc|ccc|ccc}
\toprule
\multicolumn{2}{c|}{\multirow{2}{*}{\textbf{Methods}}} & \multicolumn{3}{c|}{ESOL (1128)} & \multicolumn{3}{c|}{FreeSolv (642)} & \multicolumn{3}{c}{Lipophilicity (4200)} \\ 
\cmidrule(lr){3-5} \cmidrule(lr){6-8} \cmidrule(lr){9-11}
& & R=0.1 & R=0.2 & R=0.8 & R=0.1 & R=0.2 & R=0.8 & R=0.1 & R=0.2 & R=0.8 \\ 
\midrule
\multirow{3}{*}{\rotatebox{90}{SchNet}}
& \textit{+PCEvo }   & 1.484 & 1.568 & 1.060 & 3.490 & 2.383 & 1.958 & 1.112 & 0.854 & 0.762 \\
& \textit{Origin.}   & 1.098 & 1.046 & 0.912 & 3.972 & 4.042 & 2.255 & 1.112 & 1.067 & 0.923 \\
& \textit{Improve.} & \textcolor[rgb]{0.5,0.5,0.5}{-35.15} & \textcolor[rgb]{0.5,0.5,0.5}{-49.94} & \textcolor[rgb]{0.5,0.5,0.5}{-16.21} & \textcolor{red}{12.15} & \textcolor{red}{41.06} & \textcolor{red}{13.18} & \textcolor[rgb]{0.5,0.5,0.5}{-0.02} & \textcolor{red}{19.96} & \textcolor{red}{17.45} \\ 
\midrule

\multirow{3}{*}{\rotatebox{90}{DimeNet}}
& \textit{+PCEvo }    & 1.489 & 1.663 & 1.117 & 3.623 & 2.555 & 1.843 & 1.111 & 0.980 & 0.925 \\
& \textit{Origin.}   & 2.156 & 2.249 & 1.135 & 5.287 & 4.910 & 2.589 & 1.083 & 1.065 & 1.043 \\
& \textit{Improve.}  & \textcolor{red}{30.92} & \textcolor{red}{26.06} & \textcolor{red}{1.54} & \textcolor{red}{31.47} & \textcolor{red}{47.96} & \textcolor{red}{28.82} & \textcolor[rgb]{0.5,0.5,0.5}{-2.54} & \textcolor{red}{7.99} & \textcolor{red}{11.29} \\ 
\midrule

\multirow{3}{*}{\rotatebox{90}{ViSNet}}
& \textit{+PCEvo }    & 1.235 & 1.165 & 0.969 & 2.857 & 2.684 & 2.539 & 1.080 & 0.837 & 0.798 \\
& \textit{Origin.}  & 1.341 & 1.333 & 1.049 & 4.255 & 3.801 & 3.158 & 1.104 & 1.052 & 0.944 \\
& \textit{Improve.}  & \textcolor{red}{7.93} & \textcolor{red}{12.56} & \textcolor{red}{7.70} & \textcolor{red}{32.87} & \textcolor{red}{29.40} & \textcolor{red}{19.59} & \textcolor{red}{2.18} & \textcolor{red}{20.43} & \textcolor{red}{15.51} \\ 
\bottomrule
\end{tabular}
}
\end{table}

% % 9.9分
\section{Theoretical Analysis}
\label{sec:theory}

% This section analyzes why PCEvo improves generalization in low-data regimes by leveraging edit path decomposition. This analysis highlights two effects: edit-level supervision increases the effective sample size, and learning on edit-local inputs reduces hypothesis complexity.

This section explains why PCEvo generalizes better by decomposing sparse static endpoint regression supervision into edit-level signals along evolutionary paths.
Two mechanisms are emphasized: (i) edit decomposition increases the effective number of supervised signals; (ii) learning on edit-local representation increments reduces hypothesis complexity.

\subsection{Learning Formulation}

Given a labeled dataset $\mathcal{D}=\{(G_i,y_i)\}_{i=1}^{N}$ and the induced evolutionary dataset $\mathcal{D}_{\mathrm{evo}}$, we compare static endpoint learning and path-consistent learning.
Here $\widehat{\mathcal{R}}(\cdot)$ denotes the empirical risk (sample average of the loss) minimized during training.

\textbf{Static model.}
The static objective $\mathcal{L}_{\text{static}}$ supervises absolute property prediction on $\mathcal{D}$.
We denote by $\mathcal{H}_{\text{static}}$ the hypothesis class of all static predictors realizable by the backbone when parameters vary:
\begin{equation}
\widehat{\mathcal{R}}_{\text{static}}(f_\theta)=\frac{1}{N}\sum_{i=1}^{N}\ell\!\left(f_\theta(G_i),y_i\right),
\end{equation}
\begin{equation}
\mathcal{H}_{\text{static}}=\{\,f_\theta(\cdot)\mid \theta\in\Theta\,\}.
\end{equation}

\textbf{Path-consistent learning.}
The consistency objective $\mathcal{L}_{\text{cons}}$ enforces that the summed stepwise deltas along $\tau=(G_s^{(0)},\ldots,G_s^{(T)})$ match the endpoint difference $(y_t-y_s)$.
With $h_t=f_\theta(G_s^{(t)})$, the predicted total change is
\begin{equation}
\widehat{\Delta P}(\tau)=\sum_{t=0}^{T-1}\phi_\psi\!\left(h_{t+1}-h_t\right).
\end{equation}
We denote by $\mathcal{H}_{\text{cons}}$ the hypothesis class of all delta predictors realizable by the MLP head when its parameters vary:
\begin{equation}
\widehat{\mathcal{R}}_{\text{cons}}(\phi_\psi)=\frac{1}{|\mathcal{D}_{\mathrm{evo}}|}\sum_{\mathcal{D}_{\mathrm{evo}}}
\ell\!\left(\widehat{\Delta P}(\tau),\,y_t-y_s\right),
\end{equation}
\begin{equation}
\mathcal{H}_{\text{cons}}=\{\,\phi_\psi(\cdot)\mid \psi\in\Psi\,\}.
\end{equation}

\subsection{Generalization Bound}

We adopt Rademacher complexity to compare the generalization of static endpoint learning and edit-path learning.

\begin{assumption}[Effective sample size under mixing]
\label{assump:neff_mixing}
Edit-level samples induced by virtual paths form a weakly dependent sequence.
For instance, they satisfy $\beta$-mixing with $\sum_{k\ge 1}\beta(k)<\infty$.
Then the edit-level learning admits an effective sample size
\begin{equation}
n_{\mathrm{eff}} \ge \frac{N\bar L}{c_{\mathrm{mix}}},
\qquad
c_{\mathrm{mix}} = 1 + 2\sum_{k\ge 1}\beta(k),
\label{eq:neff}
\end{equation}
where $\bar L$ is the average path length.
\end{assumption}

Assumption~\ref{assump:neff_mixing} states that, although edit-level samples generated along paths are weakly dependent, their statistical effect can be captured by an effective sample size $n_{\mathrm{eff}}$, so the generalization bound applies by replacing $n$ with $n_{\mathrm{eff}}$.

% Then, let $\mathfrak{R}_n(\mathcal{H})$ denote the empirical Rademacher complexity.
% A standard result implies that with probability at least $1-\delta$,
% \begin{equation}
%     \mathcal{R}(f)\le \widehat{\mathcal{R}}(f)+2\mathfrak{R}_n(\mathcal{H})+3\sqrt{\frac{\log(2/\delta)}{2n}},
%     \label{eq:rad_bound}
% \end{equation}
% \noindent uniformly for all $f\in\mathcal{H}$. Under the above assumption, the same scaling is obtained for edit-level learning by replacing $n$ with $n_{\text{eff}}$ (up to mixing constants absorbed into $c_{\text{mix}}$).

Let $\mathfrak{R}_n(\mathcal{H})$ denote the empirical Rademacher complexity.
By the standard Rademacher generalization theory \cite{yin2019rademacher,bartlett2005local}, the generalization gap scales as
$O\!\big(\mathfrak{R}_n(\mathcal{H})+\sqrt{\log(1/\delta)/n}\big)$ uniformly over $f\in\mathcal{H}$.
Under Assumption~\ref{assump:neff_mixing}, edit-level learning obeys the same scaling with $n$ replaced by $n_{\mathrm{eff}}$
(up to constants absorbed into $c_{\mathrm{mix}}$).

\begin{theorem}[Edit decomposition yields a tighter generalization bound]
\label{thm:main_bound_merged}
Assume $\ell(\cdot,\cdot)$ is bounded in $[0,1]$ and is $1$-Lipschitz in its first argument.
Assume $\|h\|_2\le B_{\text{static}}$ and $\|\Delta h\|_2\le B_{\text{edit}}$.
If hypotheses in $\mathcal{H}_{\text{static}}$ and $\mathcal{H}_{\text{cons}}$ are $L$-Lipschitz with respect to their inputs, then:

\paragraph{(i) Complexity scaling.}
The empirical Rademacher complexity scales with the input radius and the sample size. In particular,
\begin{equation}
\mathfrak{R}_N(\mathcal{H}_{\text{static}})
=O\!\left(\frac{L B_{\text{static}}}{\sqrt{N}}\right),
\end{equation}
\begin{equation}
\mathfrak{R}_{n_{\text{eff}}}(\mathcal{H}_{\text{cons}})
=O\!\left(\frac{L B_{\text{edit}}}{\sqrt{n_{\text{eff}}}}\right).
\end{equation}

\paragraph{(ii) Generalization bounds.}
With probability at least $1-\delta$,
\begin{align}
\mathcal{R}_{\text{static}}
&\le \widehat{\mathcal{R}}_{\text{static}}
+ 2\mathfrak{R}_{N}(\mathcal{H}_{\text{static}})
+ 3\sqrt{\frac{\log(2/\delta)}{2N}},\\
\mathcal{R}_{\text{cons}}
&\le \widehat{\mathcal{R}}_{\text{cons}}
+ 2\mathfrak{R}_{n_{\text{eff}}}(\mathcal{H}_{\text{cons}})
+ 3\sqrt{\frac{\log(2/\delta)}{2n_{\text{eff}}}}.
\end{align}
\end{theorem}

% \paragraph{Interpretation.}
The two bounds share the same form, but differ in the effective sample size and the input radius.
Edit-path learning replaces $(N,B_{\text{static}})$ with $(n_{\text{eff}},B_{\text{edit}})$, where $n_{\text{eff}}$ is induced by path decomposition and $\Delta h$ is edit-local.
In typical settings, a single edit produces a local perturbation in representation space, so $B_{\text{edit}}\le B_{\text{static}}$.
Therefore, the edit-path bound becomes tighter whenever
\begin{equation}
\frac{B_{\text{edit}}}{\sqrt{n_{\text{eff}}}}
<
\frac{B_{\text{static}}}{\sqrt{N}}.
\label{eq:better_condition_basic}
\end{equation}

In particular, the static generalization gap scales as
\begin{equation}
\mathcal{R}_{\text{static}}-\widehat{\mathcal{R}}_{\text{static}}
=O\!\left(\frac{L B_{\text{static}}}{\sqrt{N}}+\sqrt{\frac{1}{N}}\right),
\label{eq:scaling_static}
\end{equation}
while edit-path learning follows the same scaling with $(N,B_{\text{static}})$ replaced by $(n_{\text{eff}},B_{\text{edit}})$. Finally, enforcing path consistency further restricts the admissible solutions within $\mathcal{H}_{\text{cons}}$, which can only reduce the effective complexity and tighten the above bound.

\section{Experiments}

% highest occupied molecular orbital and lowest unoccupied molecular orbital

\paragraph{Datasets.} 
We conduct our evaluations on the QM9 dataset \cite{QM9}, focusing on the prediction of the highest occupied molecular orbital ($\epsilon_{\text{HOMO}}$) and lowest unoccupied molecular orbital ($\epsilon_{\text{LUMO}}$). For physicochemical property prediction, we further employ the ESOL, FreeSolv, and Lipophilicity datasets from MoleculeNet \cite{MoleculeNet}.

% \paragraph{Baselines.} 
% To demonstrate the universality of our framework, we employed a representative set of invariant and equivariant backbones, including SchNet \cite{schnet}, DimeNet \cite{gasteiger_dimenet_2020}, Equiformer \cite{equiformer}, and ViSNet \cite{visnet}. 

\paragraph{Baselines.}
We evaluate PCEvo against a comprehensive set of backbone methods, grouped into three categories: \textbf{2D GNNs/Transformer:} GCN \cite{gcn}, GraphSAGE \cite{GraphSAGE}, GAT \cite{gat}, GIN \cite{gin}, and Moleculeformer \cite{qin2025moleculeformer}; \textbf{3D geometry-aware:} SchNet \cite{schnet}, DimeNet \cite{gasteiger_dimenet_2020}, Equiformer \cite{equiformer}, ViSNet \cite{visnet}, and GotenNet \cite{gotennet}; \textbf{Pretraining:} N-GRAM \cite{NGRAM}, GROVER \cite{GROVER}, MolCLR \cite{MolCLR},  Hu et al. \cite{Huat}, and EMPP \cite{empp}.

% These methods are well-established molecular representation models that exhibit consistently strong and stable performance across molecular property prediction tasks.

\paragraph{Implementation Details.} 
We designed specific data splitting protocols to simulate varying degrees of data scarcity. For the QM9 dataset, we constructed extremely few-shot training/testing sets containing randomly sampled 100 and 1,000 molecules. For the smaller MoleculeNet datasets, we simulated few-shot learning by using 10\%, 20\%, and 80\% of the total data for training; the 10\% data were allocated for testing.  Performance was assessed using three standard regression metrics: mean absolute error (MAE), mean squared error (MSE), and the Pearson correlation coefficient (PCC).

\subsection{Overall Experiments}

To evaluate the effectiveness and scalability of PCEvo for molecular representation, we conduct experiments on QM9 and MoleculeNet. Following our path construction procedure, for each target molecule, we retrieve the top-$N{=}5$ structural neighbors and enumerate up to $P_{\max}{=}50$ valid edit paths for each molecular pair under topological dependency constraints.

On QM9, we adopt two low-resource settings with $N{=}100$ and $N{=}1000$ labeled training samples (results in Table~\ref{tab:qm9_results}). PCEvo yields consistent improvements across four 3D backbones (SchNet, DimeNet, Equiformer, and ViSNet), and the correlation recovery is particularly pronounced in few-shot settings ($N{=}100$): for $\epsilon_{\mathrm{HOMO}}$, Equiformer improves PCC from 0.07 to 0.53 (+0.46). Moreover, compared with recent strong baselines such as EMPP, Moleculeformer, and GotenNet, PCEvo achieves better performance under the same few-shot protocol. These results indicate that when the chemical space is sparsely covered, PCEvo can effectively enhance the generalization of backbone models.

On MoleculeNet, since the benchmarks are intrinsically small-scale, we evaluate PCEvo under both the standard split (80\% training; Table~\ref{tab:MoleculeNetStandard}) and explicit low-resource splits with training ratios $R\in{0.1,0.2,0.8}$ (Table~\ref{tab:MoleculeNetFewShot}). The results show that integrating PCEvo into SchNet, DimeNet, and ViSNet leads to consistent gains across tasks and training ratios. The improvement on FreeSolv is especially evident; for example, with SchNet at $R{=}0.2$, RMSE decreases from 4.042 to 2.383. Meanwhile, under the standard split, PCEvo also outperforms several widely used pretraining baselines: for instance, $\mathrm{PCEvo}_{\;\text{SchNet}}$ achieves an average RMSE of 1.260, improving over MolCLR (1.401) by 0.141.

% This suggests that even under the full-data setting, PCEvo can consistently improve different backbones and achieve better overall performance than multiple pretraining-based methods.

Overall, PCEvo exhibits stable and transferable improvements across backbones and few-shot settings, and compares favorably to a range of pretraining methods. These results suggest that moving beyond static endpoint regression by decomposing supervision into edit-wise steps and enforcing path-consistency can yield more robust molecular representations in few-shot settings.

\subsection{Ablation Study}
\label{sec:ablation_cons}

% \begin{table}[t!]
% \centering
% \caption{Ablation on the path-consistency constraint $\mathcal{L}_{\text{cons}}$.}
% \label{tab:ablation_cons}
% \small
% \setlength{\tabcolsep}{6pt}
% \begin{tabular}{c c c c c c}
% \toprule
% $P_{\max}$ & $N$ & $\mathcal{L}_{\text{cons}}$ & MAE $\downarrow$ & MSE $\downarrow$ & PCC $\uparrow$ \\
% \midrule
% 5 & 1 & \cmark & 0.487 & 0.430 & 0.145 \\
% 5 & 1 & \xmark & 0.672 & 0.881 & -0.0883 \\
% 5 & 3 & \cmark & 0.385 & 0.265 & 0.659 \\
% 5 & 3 & \xmark & 0.518 & 0.480 & -0.108 \\
% 5 & 5 & \cmark & \textbf{0.319} & \textbf{0.179} & \textbf{0.761} \\
% 5 & 5 & \xmark & 0.390 & 0.273 & 0.639 \\
% \midrule
% 1 & 1 & \cmark & 0.687 & 0.942 & -0.0859 \\
% 1 & 1 & \xmark & 0.671 & 0.885 & -0.0886 \\
% 1 & 3 & \cmark & 0.672 & 0.894 & -0.0889 \\
% 1 & 3 & \xmark & 0.677 & 0.904 & -0.0866 \\
% 1 & 5 & \cmark & 0.494 & 0.443 & 0.127 \\
% 1 & 5 & \xmark & 0.486 & 0.428 & 0.153 \\
% \bottomrule
% \end{tabular}
% \end{table}

\begin{wraptable}{r}{0.6\columnwidth}
\vspace{-2.5em}
\caption{Ablation study.}
\label{tab:ablation_cons}
\footnotesize
\setlength{\tabcolsep}{2.5pt}
\centering
\begin{tabular}{c c c r r r}
\toprule
$P_{\max}$ & $N$ & $\mathcal{L}_{\text{cons}}$ & MAE$\downarrow$ & MSE$\downarrow$ & PCC$\uparrow$ \\
\midrule
$^{\phantom{*}}$5 & 1 & \cmark & 0.487 & 0.430 & 0.145 \\
$^{\phantom{*}}$5 & 1 & \xmark & 0.672 & 0.881 & -0.088 \\
$^{\phantom{*}}$5 & 3 & \cmark & 0.385 & 0.265 & 0.659 \\
$^{\phantom{*}}$5 & 3 & \xmark & 0.518 & 0.480 & -0.108 \\
$^{\phantom{*}}$5 & 5 & \cmark & \textbf{0.319} & \textbf{0.179} & \textbf{0.761} \\
$^{\phantom{*}}$5 & 5 & \xmark & 0.390 & 0.273 & 0.639 \\
\midrule
$^{\phantom{*}}$1 & 1 & \cmark & 0.687 & 0.942 & -0.085 \\
*1 & 1 & \xmark & 0.469 & 0.492 & 0.341 \\
$^{\phantom{*}}$1 & 3 & \cmark & 0.672 & 0.894 & -0.088 \\
$^{\phantom{*}}$1 & 3 & \xmark & 0.677 & 0.904 & -0.086 \\
$^{\phantom{*}}$1 & 5 & \cmark & 0.494 & 0.443 & 0.127 \\
$^{\phantom{*}}$1 & 5 & \xmark & 0.486 & 0.428 & 0.153 \\
\bottomrule
\end{tabular}

% \begin{flushleft}
% \footnotesize $^{*}$ denotes the vanilla SchNet results.
% \end{flushleft}

\noindent\makebox[0.6\columnwidth][l]{\footnotesize $^{*}$ denotes the vanilla SchNet results.}

\vspace{-1em}
\end{wraptable}

% We conduct a strict ablation to isolate the effect of the path-consistency constraint $\mathcal{L}_{\text{cons}}$. We use SchNet as the backbone, train under the 100-shot setting, and evaluate on a disjoint 100 molecules as validation set with $\epsilon_{\text{HOMO}}$ as the target property. Across all runs, we keep the static endpoint supervision $\mathcal{L}_{\text{static}}$ enabled, and control the training data construction by fixing the same neighbor budget $N$ and the same per-pair path budget $P_{\max}$. Therefore, within each $(N, P_{\max})$ setting, the only change is whether $\mathcal{L}_{\text{cons}}$ is activated.

To isolate the effect of the path-consistency constraint $\mathcal{L}_{\text{cons}}$, we conduct a controlled ablation study. We use SchNet as the backbone, train under the 100-shot, and evaluate on a held-out validation set of 100 disjoint molecules, with $\epsilon_{\text{HOMO}}$ as the target property. Across all runs, we keep the static endpoint supervision $\mathcal{L}_{\text{static}}$ enabled. Therefore, within each $(N, P_{\max})$ configuration, the only difference is whether $\mathcal{L}_{\text{cons}}$ is enabled.

\paragraph{Main finding: $\mathcal{L}_{\text{cons}}$ is effective when multiple paths are available.} Table~\ref{tab:ablation_cons} shows that relative to the single path setting ($P_{\max}=1$), multi-path training ($P_{\max}=5$) consistently improves performance when the path-consistency constraint $\mathcal{L}_{\text{cons}}$ is enabled. This trend holds across all neighbor budgets $N\in{1,3,5}$, indicating that the benefit of $\mathcal{L}_{\text{cons}}$ is robust to the choice of $N$. In contrast, simply increasing the number of paths without $\mathcal{L}_{\text{cons}}$ does not lead to reliable gains and can even degrade performance in few-shot settings.

% Table~\ref{tab:ablation_cons} shows that under multi-path training ($P_{\max}=5$), enabling $\mathcal{L}_{\text{cons}}$ yields consistent improvements across all neighbor budgets. This indicates that the constraint provides a strong regularization signal when the model can observe multiple valid decompositions between the same endpoints.

\paragraph{On the role of naive duplication.}
In our training construction, endpoint molecules are reused across different molecular pairs and paths. Such reuse mainly increases the number of training instances, but does not necessarily introduce additional chemical diversity at the endpoints; therefore, the observed gains cannot be explained by sample count alone. Table~\ref{tab:ablation_cons} provides a controlled comparison supporting this point: once the path-consistency constraint is removed, increasing the path budget ($P_{\max}>1$) or the neighbor budget ($N>1$) no longer yields consistent improvements; instead, performance becomes unstable under some configurations.

\subsection{Hyperparameter Sensitivity Analysis}

\begin{figure}[t!]
    \centering
    \includegraphics[width=\columnwidth]{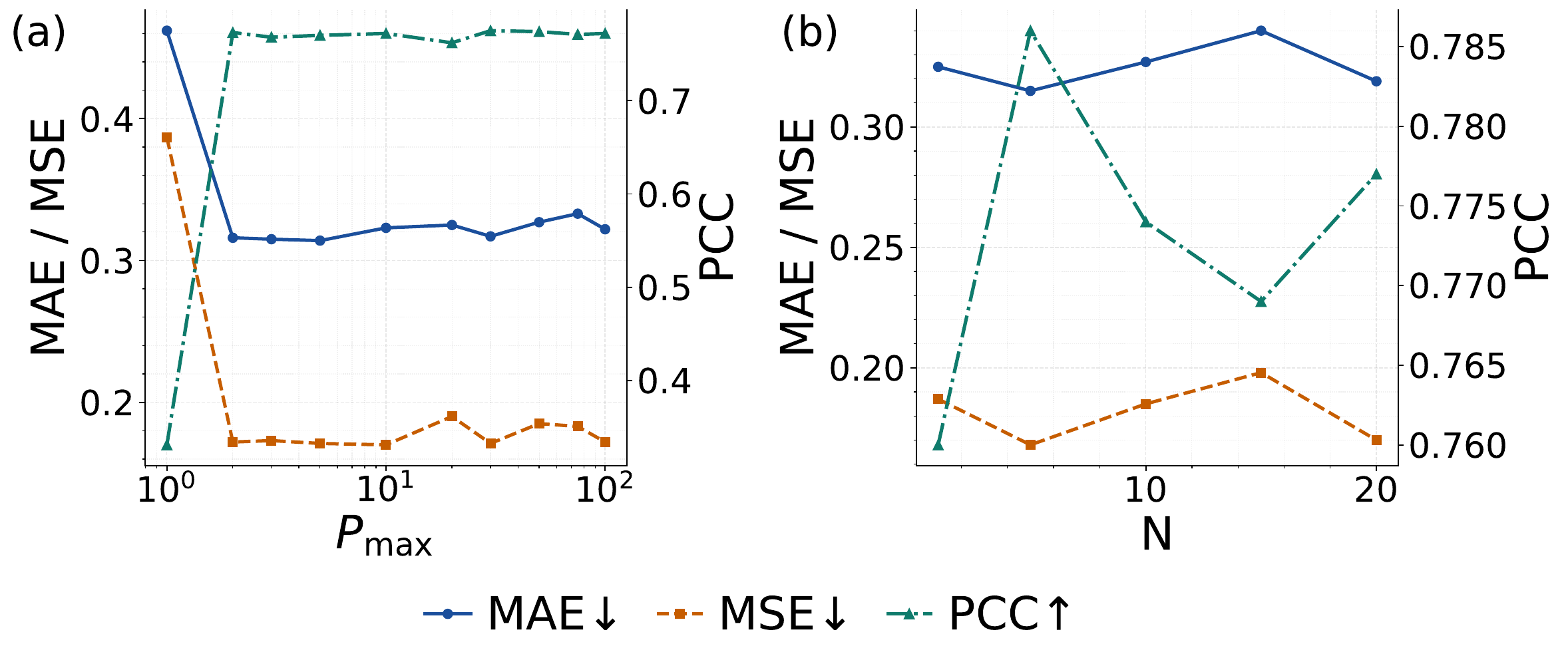}
    \caption{Hyperparameter sensitivity analysis on QM9 with SchNet under the 100-shot setting. (a) fixes $N{=}10$ and varies $P_{\max}$, (b) fixes $P_{\max}{=}50$ and varies $N$.}
    \label{fig:hparam_sensitivity}
\end{figure}

To analyze the impact of the maximum number of virtual paths $P_{\max}$ and the number of source neighbors $N$ on PCEvo, we conduct experiments on the QM9 dataset under the 100-shot setting using SchNet as the backbone, with results shown in Fig.~\ref{fig:hparam_sensitivity}. Regarding the number of virtual paths, we observe that a single path ($P_{max}=1$) yields negligible gains over the baseline, indicating that simple augmentation is insufficient. However, increasing $P_{max}$ significantly boosts performance by enforcing consistency across diverse topological orderings. This confirms that path-consistent via virtual evolutionary is central to the PCEvo's success. Regarding the context size $N$, a clear trade-off exists between diversity and relevance. Performance peaks at $N=5$; utilizing larger neighborhoods ($N \ge 15$) degrades accuracy, as distant neighbors introduce chemically irrelevant scaffolds that generate noisy and overly complex edit paths.

\section{Conclusion}
In this work, we propose PCEvo, a path-consistent molecular representation method that augments conventional static regression supervision with a virtual evolutionary edit-path consistency constraint. This design enables the model to learn how local structural edits accumulate into global property variations, thereby yielding more robust structure--property mappings. Experiments on QM9 and MoleculeNet under few-shot settings show that PCEvo consistently reduces error, improves stability, and achieves SOTA performance against strong baselines.

\bibliographystyle{named}
\bibliography{ijcai26}

\end{document}